\newcommand{\text}[1]{{\rm{#1}}}
\newcommand{\textsubscript}[1]{{$_{#1}$}}
\newcommand{\Fig}{{Figure}}

\documentclass[showkeys,showpacs,prb,twocolumn, superscriptaddress,bibnotes]{revtex4}
\usepackage{color}
\usepackage{graphicx}

\makeatletter

\begin{document}
\title{First-principles study of the thermoelectric properties of strained graphene nanoribbons}

\author{Pei Shan Emmeline Yeo}
\affiliation
{Institute of High Performance Computing, Agency
for Science, Technology and Research, 1 Fusionopolis Way, \#16-16
Connexis, Singapore 138632}
\affiliation
{Department of Chemistry, National University of Singapore, 
3 Science Drive 3, Singapore 117543 }
\author{Michael B. Sullivan}
\affiliation
{Institute of High Performance Computing, Agency
for Science, Technology and Research, 1 Fusionopolis Way, \#16-16
Connexis, Singapore 138632}
\author{Kian Ping Loh}
\affiliation
{Department of Chemistry, National University of Singapore, 
3 Science Drive 3, Singapore 117543 }
\author{Chee Kwan Gan}
\email{ganck@ihpc.a-star.edu.sg}
\affiliation
{Institute of High Performance Computing, Agency
for Science, Technology and Research, 1 Fusionopolis Way, \#16-16
Connexis, Singapore 138632}

\begin{abstract}

We study the transport properties, in particular, the thermoelectric
figure of merit (\emph{$ZT$}) of armchair graphene nanoribbons, AGNR-$N$ (for $N=4-12$, with widths
ranging from $3.7$ to $13.6$~\AA) through
strain engineering, where $N$ is the number of carbon dimer lines
across the AGNR width. We find that the tensile strain applied to
AGNR-$N$ changes the transport properties by modifying the electronic
structures and phonon dispersion relations. The tensile strain increases
the $ZT$ value of the AGNR-$N$ families with $N=3p$ and $N=3p+2$,
where $p$ is an integer. Our analysis based on accurate density-functional
theory calculations suggests a possible route to increase the $ZT$
values of AGNR-$N$ for potential thermoelectric applications.

\end{abstract}
\maketitle

\section{Introduction}

Currently thermoelectric materials receive considerable attention\cite{Szczech2011}
due to their ability to produce electricity from waste heat generated
in, for example, power plants and refrigeration units. The efficiency
of a thermoelectric material is characterized by the figure of merit

\[
ZT=\frac{G_{{\rm e}}S^{2}}{K_{{\rm e}}+K_{{\rm ph}}}T
\]
where $G_{\text{e}}$ is the electrical conductance, $S$ is the Seebeck
coefficient, $K_{\text{e}}$ ($K_{\text{ph}}$) is the thermal conductance
due to electrons (phonons), and $T$ is the absolute temperature.
It is challenging to engineer thermoelectric materials because the
parameters $G_{\text{e}}$, $S$, $K_{\text{e}}$, and $K_{\text{ph}}$
are intricately interrelated; an attempt to improve one parameter
usually detrimentally affects the others.\cite{Snyder2008} It is
generally agreed that for thermoelectric generators to be viable,
a material with $ZT\sim2-4$ is required.\cite{Sootsman2009,Tritt2011}

Current state-of-the-art thermoelectric materials\cite{Tritt2011,Teweldebrhan2010,Goyal2010,Teweldebrhan2010a}
such as single-layer Bi\textsubscript{2}Te\textsubscript{3} and
AgPb\textsubscript{18}SbTe\textsubscript{20} possess $ZT$ between
$2$ and $3$, but they are composed of high atomic number elements,
thus making them both expensive and heavy. Graphene, which is composed
of a hexagonal network of lightweight carbon atoms, with extremely
high electron mobility and long electron mean
free paths,\cite{Bolotin2008} is a potential thermoelectric material.
Experimental measurements of $S$ for graphene showed values of $80$~$\mu$VK\textsuperscript{-1}
at $300$~K,\cite{Zuev2009} $39$~$\mu$VK\textsuperscript{-1}
at $255$~K,\cite{Wei2009} and $100$~$\mu$VK\textsuperscript{-1}
at $280$~K,\cite{Checkelsky2009} which are moderate compared to
$150-850$~$\mu$VK\textsuperscript{-1} at room temperature
for the inorganic materials,\cite{Harman2002,Hsu2004,Ohta2007} but
comparable to other organic thermoelectric materials such as conducting
polymers.\cite{Dubey2011}

Since graphene has very high $K_{\text{ph}}$,\cite{Balandin2008,Faugeras2010,Balandin2011,Nika2012}
a common approach to increase $ZT$ of graphene-related materials
is to reduce $K_{\text{ph}}$. For example, edge disorder decreases
the phonon mean free path and therefore reduces $K_{\text{ph}}$,
which may increase $ZT$. However, edge disorder impacts the electronic
structure of the materials as well. In the case of armchair graphene nanoribbons (GNRs),
\cite{Ouyang2009,Mazzamuto2012}
edge disorder turns out to be detrimental to $ZT$; whereas for zigzag
graphene nanoribbons, a high $ZT\sim4$ was obtained.\cite{Sevincli2010}

Introducing vacancies into GNRs may also reduce phonon thermal conductance.
Randomly distributed vacancies tend to decrease $ZT$,\cite{Ouyang2009,Mazzamuto2012}
while periodically distributed lattice defects increases $ZT$. The
maximal $ZT$ attainable\cite{Gunst2011,Zhao2011a} is $\sim0.2$ to $0.3$.

Other methods to suppress $K_{\text{ph}}$ involves crafting graphene
into novel nanostructures. Graphene nanojunctions, which consist of
graphene domains with different widths connected together,\cite{Chen2010}
demonstrated a maximum $ZT\sim0.6$. In a similar vein, attaching ``stub''
structures to the edges of GNRs\cite{Xie2012} resulted in $ZT\sim0.25$.
In the case of kinked GNRs, a maximum $ZT\sim0.4$ can be achieved.\cite{Huang2011}
Cutting graphene into ``nanowiggles''\cite{Liang2012} delivered
a maximum $ZT\sim0.79$, while crafting graphene into structures with
alternating armchair-edge and zigzag-edge domains\cite{Mazzamuto2011} delivered $ZT\sim1$.

There have also been attempts to combine the two methods above by
etching periodic vacancies into novel graphene nanostructures. A maximum
$ZT$ of $0.4$ and even $5$ have been reported in these structures.\cite{Mazzamuto2012,Chang2012}

Finally, the thermoelectric properties of graphene may be improved
by incorporating heteroatoms\cite{Ni2009,Yang2012} or isotopes\cite{Chen2012,Balandin2012} into it.
For example, by attaching hydrogen
atoms on the surface of the GNRs, $ZT\sim26$ was reported.\cite{Ni2009}
The thermoelectric properties of hybrid nanoribbons consisting of
alternating graphene and hexagonal boron nitride regions have also
been investigated.\cite{Yang2012} A maximum $ZT\sim0.7$ was observed.

The strategies mentioned above face several challenges as they involve
engineering complex shapes out of GNRs. Furthermore, foreign entities
incorporated into graphene are either removed,\cite{Elias2009} or
nucleate to form large clusters,\cite{Ci2010} when subjected to elevated
temperatures. Inspired by previous works that showed that tensile strain
reduces the $K_{\text{ph}}$ of graphene-related materials,\cite{Zhai2011,Wei2011,Yeo2012}
we examine the effect of strain on the $ZT$ value of armchair graphene nanoribbons (AGNRs). Compared
to the methods mentioned above, tensile strain\cite{Lu2012} is relatively easier
to be imposed on AGNRs, thus enabling a possible route to manipulate
the $ZT$ values.

\section{Methodology}

To calculate the thermoelectric properties of the AGNR-$N$ with a
strain parameter $\varepsilon$, we use the Landauer approach to calculate
the transport properties, where the electrical conductance $G_{\text{e}}$,
the Seebeck coefficient $S$, and the thermal conductance due to electrons
$K_{\text{e}}$ are obtained in the linear response regime under an open
circuit condition:

\begin{equation}
G_{\text{e}}(\varepsilon,T)=e^{2}L_{0}\text{,}
\label{eq:G}
\end{equation}

\begin{equation}
S(\varepsilon,T)=-\frac{1}{eT}\frac{L_{1}}{L_{0}}\text{,}
\end{equation}

\begin{equation}
K_{\text{e}}(\varepsilon,T)=\frac{1}{T}\left(L_{2}-\frac{L_{1}^{2}}{L_{0}}\right)\text{,}
\end{equation}
where the $n$th order Lorenz function is given by

\begin{equation}
L_{n}=\frac{2}{h}\int_{-\infty}^{\infty}\theta_{\text{e}}(E)(E-\mu)^{n}\left(-\frac{\text{\ensuremath{\partial}}f(E,\mu,T)}{\text{\ensuremath{\partial}}E}\right)\text{d}E\text{.}
\label{eq:L}
\end{equation}

In the above equations, $e$ is the elementary charge, $f(E,\mu,T)=[\text{e}^{(E-\mu)/kT}+1]^{-1}$
is the Fermi-Dirac distribution with energy $E$, chemical potential
$\mu$, and temperature $T$. The Planck and Boltzmann constants are
$h$ and $k$, respectively. $\theta_{\text{e}}(E)$ is the electronic
transmission function, which is the number of effective modes available
for electronic transport at $E$. We expect
Eq.~\ref{eq:G} to Eq.~\ref{eq:L} to be valid even for the case
of one-atom thick AGNRs since the approximations made to derive them
do not take the dimensionality of the system into account. Assuming ballistic transport 
(since our system sizes are much smaller than the typical electron\cite{Bolotin2008a} and 
phonon\cite{Ghosh2008} mean free paths)
and
completely uniform contact and transport regions, we can calculate
$\theta_{\text{e}}(E)$ by counting the number of bands at $E$ from
the electronic band structure along the transport direction of interest.\cite{Markussen2008,Jeong2010}
We note here that for more general cases, $\theta_{\text{e}}(E)$
may be calculated using the nonequilibrium Green's function method.\cite{Wang2013}
Since the highest doping concentration achievable\cite{Lee2010} so far by molecular
charge-transfer and hetero-atom doping of graphene 
is $\sim10^{13}$~cm\textsuperscript{-2},
we investigate doping concentrations in AGNR-$N$ from $-10^{14}$
to $10^{14}$~cm\textsuperscript{-2}; with a negative (positive)
concentration representing electron or n-type doping (hole or p-type
doping). We consider both n- and p-type doping because thermoelectric
devices requires both types of materials. We restrict our study to
temperatures from $200$ to $800$~K.

\begin{figure}
\begin{centering}
\includegraphics[width=8cm]{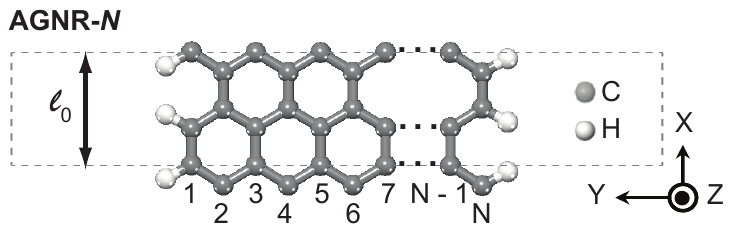} 
\par\end{centering}

\caption{The unit cell of AGNR-$N$, with optimized length $\ell_{0}$. The
dashed grey lines denote the boundaries of the unit cell.}

\label{Fig:unitcell} 
\end{figure}

We perform nonspin-polarized density-functional theory (DFT) calculations
on AGNRs using the SIESTA package.\cite{Soler2002a} The unit cell
of AGNR-$N$ is shown in \Fig~\ref{Fig:unitcell}, where $N$ is
the number of carbon dimer lines across the AGNR, and each carbon
atom at the edge is terminated with a single hydrogen atom. A vacuum
separation of at least 15~\AA{}\ is imposed in the $y$ and $z$
directions, where we use the convention adopted in \Fig\ref{Fig:unitcell}.
The local density approximation (LDA) is used for the exchange\textendash{}correlation
functional. Troullier--Martins pseudopotentials and double-$\zeta$
basis sets are used for the carbon and hydrogen atoms. A mesh cutoff
of 400~Ryd is used. We obtain the optimized length $\ell_{0}$ of
each $N$ in the $x$ direction (i.e., the transport direction) by
relaxing the atomic positions of AGNR-$N$ with different ribbon lengths
$\ell$. The atomic relaxation is performed using the conjugate gradient
algorithm with a force tolerance criterion of $10^{-3}$~eV/\AA{}.
The total energies of the relaxed structures are fitted to a polynomial
function as a function of $\ell$ to obtain $\ell_{0}$. For the strained
AGNRs, we use unit cells with $\ell=(1+\varepsilon)\ell_{0}$, with
different strain parameter $\varepsilon$ values of $0.025$, $0.050$,
$0.075$, and $0.100$. We note that
AGNRs have compressive edge stresses that tend to cause them 
to buckle.\cite{Gan2010,Bao2009,Kumar2010} Since a large unit cell of a buckled AGNR imposes
a huge computational demand on accurate DFT calculations,
we consider the AGNRs under tensile strain in this study.

The thermal conductance due to lattice vibrations $K_{{\rm ph}}$
is calculated according to 
\[
K_{\text{ph}}(\varepsilon,T)=\int_{0}^{\infty}h\nu\theta_{\text{ph}}(\nu)\frac{\text{\ensuremath{\partial}}n_{\text{B}}(\nu,T)}{\text{\ensuremath{\partial}}T}\text{d}\nu\text{,}
\]
where $n_{\text{B}}(\nu,T)=(\text{e}^{h\nu/kT}-1)^{-1}$ is the
Bose-Einstein distribution with frequency $\nu$, and $\theta_{\text{ph}}(\nu)$
is the phonon transmission function obtained using the counting method
on the phonon dispersion relations. The supercell force-constant method
is used to perform the phonon calculations.\cite{Yeo2012,Gan2006,Zhao11v84}

\section{Results and Discussion}

\begin{figure}
\begin{centering}
\includegraphics[width=7.2cm]{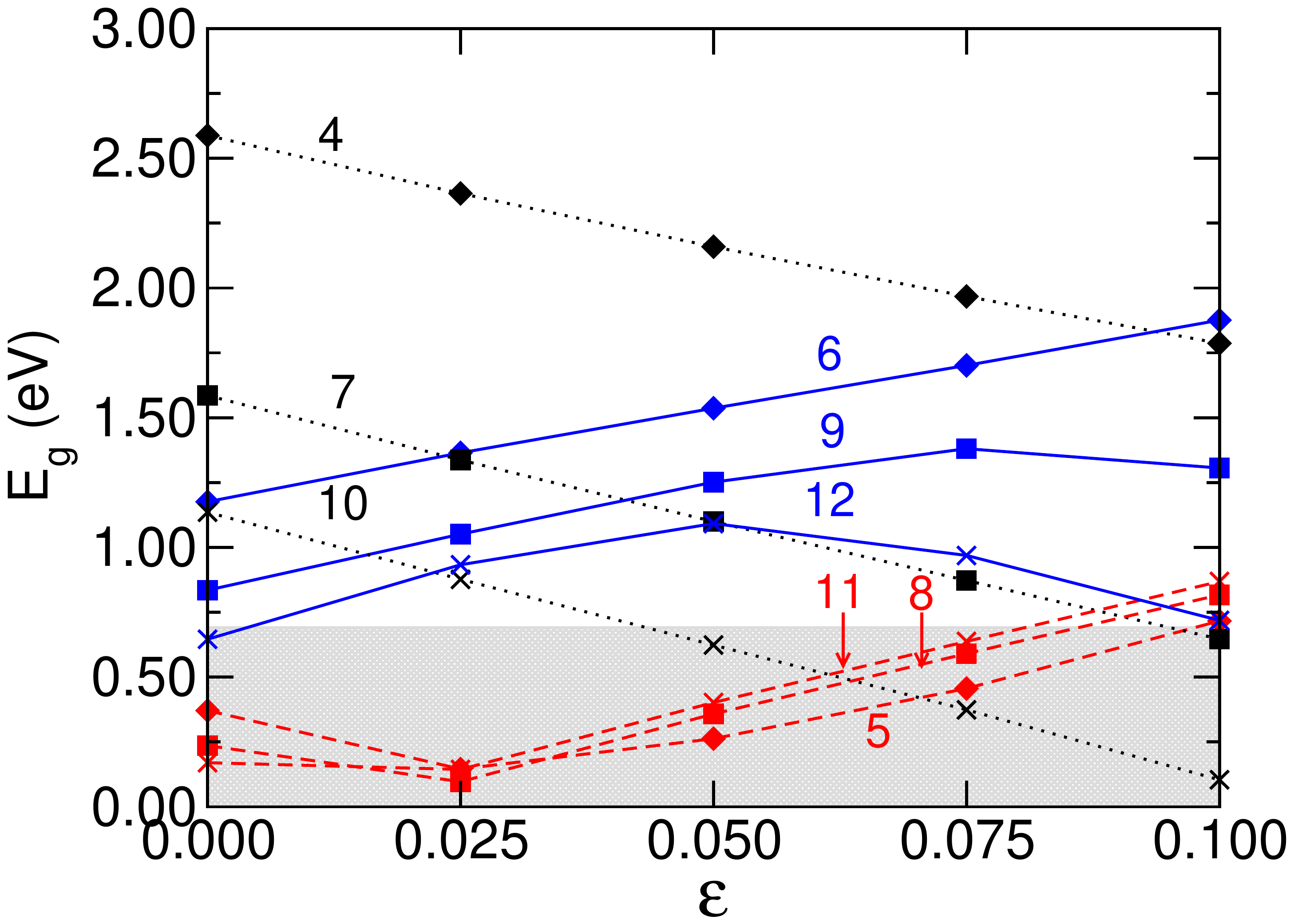} 
\par\end{centering}

\caption{Electronic band gap $E_{{\rm g}}$ versus strain parameter $\varepsilon$
for AGNR-$N$. The numbers within the graph represent $N$. The
$N=3p+1$, $N=3p$, and $N=3p+2$ families are represented by dotted lines (black),
solid lines (blue), and dashed lines (red), respectively. The grey region indicates $E_{g}<10kT$
for $T=800$~K, where bipolar transport is important.}

\label{Fig:E_g} 
\end{figure}

We first examine the effect of tensile strain on the electronic band
structures of the AGNR-$N$. The electronic band gap $E_{{\rm g}}$
has a very important influence on the thermoelectric behavior of materials:
we need an $E_{g}$ of at least $6kT$ to $10kT$ to prevent bipolar
transport because concurrent electron and hole transport leads to
an opposing effect that reduces the Seebeck coefficient. \Fig~\ref{Fig:E_g}
shows the band gap $E_{{\rm g}}(\varepsilon)$ of the AGNRs as a function
of $\varepsilon$. It is well-known that the use of the local density
approximation for the exchange-correlation functional underestimates
$E_{{\rm g}}$, but we expect the correct overall qualitative trends
to be obtained.\cite{Yang2007} In our calculations using LDA, 
$E_{{\rm g}}$ ranges from $2.59$~eV for AGNR-$4$ to $0.65$~eV for 
AGNR-$12$. In comparison, with GW calculations, $E_{{\rm g}}$
varies from $5.56$~eV for AGNR-$4$ to $1.67$~eV for AGNR-$12$.\cite{Yang2007}
 The $E_{{\rm g}}$ of AGNR-$N$ for $\varepsilon=0.00$
depends on $N$ through a 3-family behavior: $E_{{\rm g}}(N=3p+1)>E_{{\rm g}}(N=3p)>E_{{\rm g}}(N=3p+2)$,
where $p$ is an integer.\cite{Son2006a} \Fig~\ref{Fig:E_g} shows
how $E_{{\rm g}}$ varies with $\varepsilon$ for different families.
For the family with the largest $E_{{\rm g}}$, $N=3p+1$, $E_{{\rm g}}$
decreases linearly with $\varepsilon$. For the family of $N=3p$,
$E_{{\rm g}}$ increases with $\varepsilon$ except for large $\varepsilon$
for $N=9$ and $12$. For the family of $N=3p+2$, $E_{{\rm g}}$
generally increases with $\varepsilon$ for all $N$, except at $\varepsilon=0.025$.
The maximum percentage change to $E_{{\rm g}}$ for $\varepsilon=0.0-0.1$
is substantial as it ranges from $\sim-30$\% in AGNR-$4$ to $\sim+400$\%
in AGNR-$11$. For much wider AGNRs ($N>12$), a previous study\cite{Sun2008} noted
that $E_{{\rm g}}$ shows a zigzag fluctuation with $\varepsilon$.
This suggests that it is much harder to tune $E_{{\rm g}}$ by modifying
$\varepsilon$ for large $N$.

\begin{figure}
\begin{centering}
\includegraphics[width=8cm]{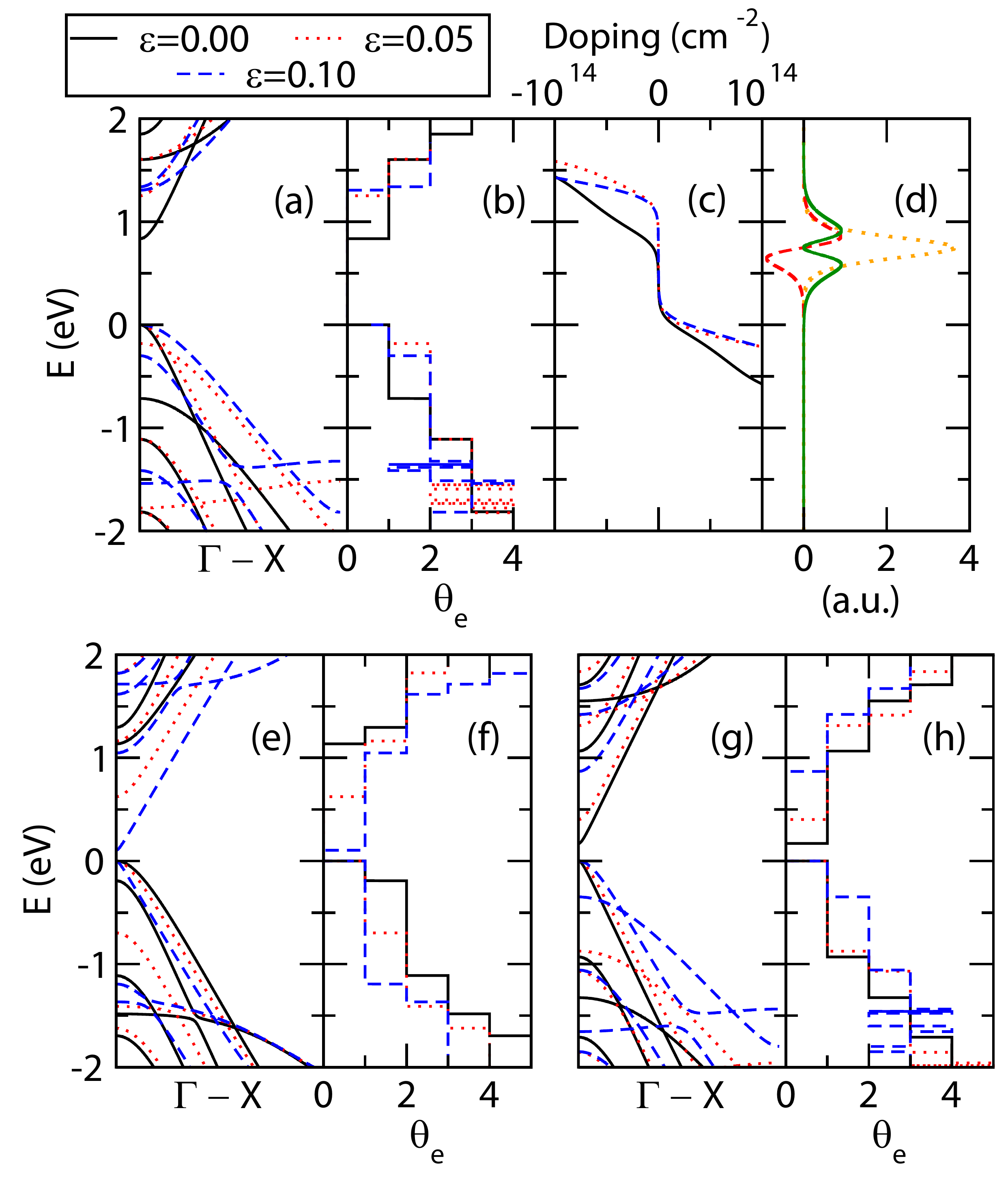} 
\par\end{centering}

\caption{The electronic band structure and corresponding $\theta_{\text{e}}(E)$
of AGNR-$9$ (a, b), AGNR-$10$ (e, f), and AGNR-$11$(g, h). (c)
shows the variation of $\mu$ for AGNR-$9$ with doping concentration
(negative concentration represents n-type doping) at $T=800$~K.
(d) shows the values of $-(E-\mu)^{n}\frac{\text{\ensuremath{\partial}}f}{\text{\ensuremath{\partial}}E}$
as a function of $E$ for $n=0$ (dotted orange line), $n=1$ (dashed red line), and $n=2$
(solid green line), for the determination of $L_{n}$. The $\mu$ is chosen such
that it maximizes $ZT$ for AGNR-$9$ under n-type doping at $T=800$~K
and $\varepsilon=0.00$. The valence band maximum is set to $0$~eV
in all figures. }

\label{Fig:bandstruc} 
\end{figure}
The Lorenz functions in Eq.~\ref{eq:L} for the evaluation of
$G_{{\rm e}}$, $|S|$, and $K_{{\rm e}}$ depend critically on the
location of $\mu$ and the extent (or spread) of $\frac{-\text{\ensuremath{\partial}}f}{\text{\ensuremath{\partial}}E}$,
both of which are controlled by the doping concentration and temperature.
The values of $\mu$ for AGNR-$9$ as a function of doping concentration
for different $\varepsilon$ are shown in \Fig~\ref{Fig:bandstruc}(c).
The spread of $\frac{-\text{\ensuremath{\partial}}f}{\text{\ensuremath{\partial}}E}$
spans only $\sim10kT$ around $\mu$ (see \Fig~\ref{Fig:bandstruc}(d)),
and thus severely restricts the range of $E$ for the integration of
$L_{n}$. As a reference, we note that at $T=800$~K, $10kT$ corresponds
to $0.69$~eV. We shall therefore study the changes of $\theta_{{\rm e}}(E)$
around the conduction (valence) band edge induced by $\varepsilon$
for n-type (p-type) doping to obtain a qualitative understanding of
the variations of $|S|$, $G_{{\rm e}}$, and $K_{{\rm e}}$ as a
result of $\varepsilon$ and temperature.

\begin{figure*}[t]
\begin{centering}
\includegraphics[width=15cm]{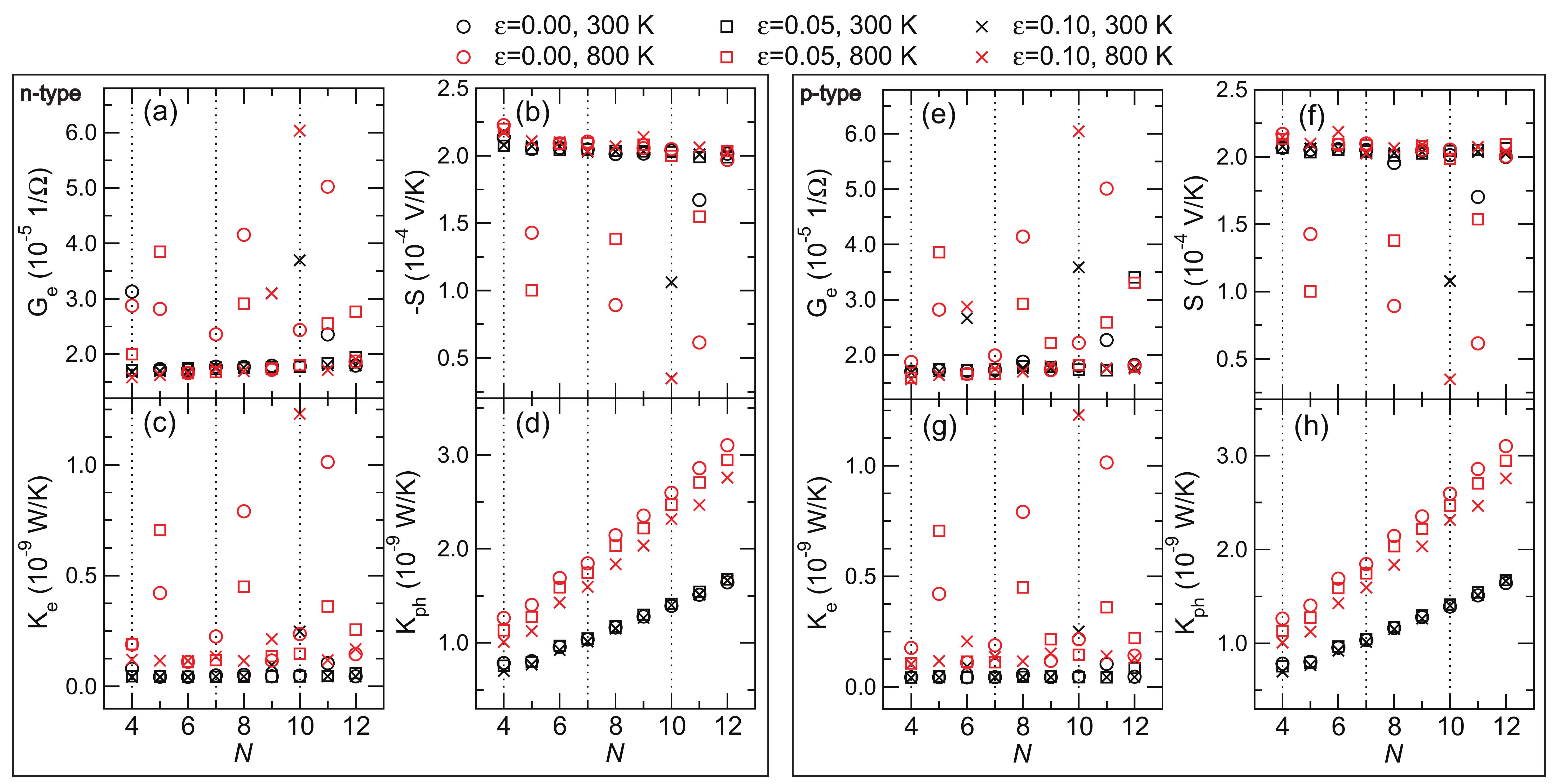} 
\par\end{centering}

\caption{Values of $G_{\text{e}}$, $|S|$, $K_{\text{e}}$, and $K_{\text{ph}}$
for $\varepsilon=0.00$, $0.05$, and $0.10$ at $T=300$ and $800$~K.
n-type [(a) to (d)] doped and p-type [(e) to (h)]
doped AGNRs are considered. $\mu$ are chosen
to maximize the $ZT$ values. The vertical dotted grey lines identify
the AGNRs belonging to the $N=3p+1$ family. }

\label{Fig:N_vs_ABS_ZT_params} 
\end{figure*}

\Fig~\ref{Fig:bandstruc}(a, e, g) shows the electronic band structures,
and \ref{Fig:bandstruc}(b, f, h) the $\theta_{\text{e}}(E)$ of the
representative AGNR-$N$ from each family, $N=9$, $10$, and $11$.
From \Fig~\ref{Fig:bandstruc}(a), we observe that AGNR-$9$ from
the $N=3p$ family displays an increase in $\theta_{\text{e}}(E)$
near the band edges for non-zero $\varepsilon$. 
From \Fig~\ref{Fig:bandstruc}(f),
AGNR-$10$ (a member of the $N=3p+1$ family) shows a decrease in
$\theta_{\text{e}}(E)$ around the band edges. Finally, members of
the $N=3p+2$ family display very little changes in $\theta_{\text{e}}(E)$
to within $\sim 0.69$~eV (the range at which $\frac{-\partial f}{\partial E}$
is nonzero at $800$~K) around the band edges with increasing $\varepsilon$;
this is shown in \Fig~\ref{Fig:bandstruc}(h) for AGNR-$11$.

We plot the various transport parameters for the determination of
$ZT$ for n- and p-type doping in \Fig~\ref{Fig:N_vs_ABS_ZT_params}
at $300$ and $800$~K with corresponding $kT$ values of $0.026$
and $0.069$~eV, respectively. At a moderate temperature such as
$300$~K, the transport is governed by monopolar transport, where
the values of $G_{{\rm e}}$, $|S|$, and $K_{{\rm e}}$ vary only
slightly with changes with $\varepsilon$. Physically this means electronic
excitation is limited at this temperature. The only exception is AGNR-$10$
at $\epsilon=0.100$, where its $E_{g}$ is so small compared to $10kT$
that bipolar transport governs and causes drastic changes in $G_{{\rm e}}$,
$|S|$, and $K_{{\rm e}}$. At a high temperature $T=800$~K, we
expect AGNR-$N$ with $E_{g}\le10kT=0.69$~eV to be affected by bipolar
transport, where $G_{{\rm e}}$ and $K_{{\rm e}}$ will generally
increase due to the presence of both types of carriers (i.e., electrons
and holes) for charge and heat transport, but $|S|$ will decrease due to the
opposing effect we mentioned earlier. This is generally evident in
\Fig~\ref{Fig:N_vs_ABS_ZT_params} where the values of $G_{e}$,
$|S|$, and $K_{e}$ change drastically at $T=800$~K compared to
that for $T=300$~K for all strain values, as long as $E_{g}(\epsilon)\le0.69$~eV.

\begin{figure}
\begin{centering}
\includegraphics[width=7.2cm]{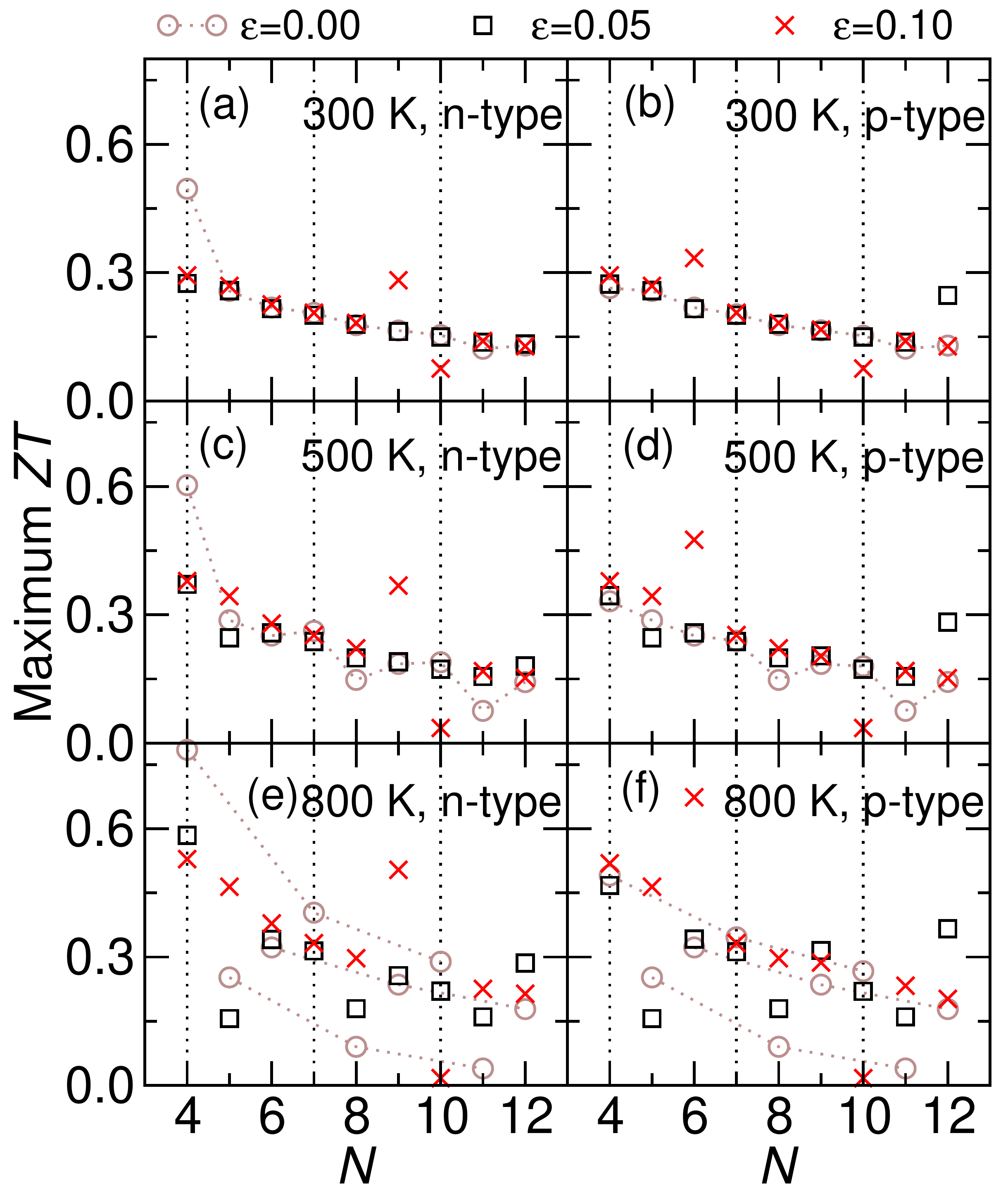} 
\par\end{centering}

\caption{Maximum $ZT$ attainable at $T=300$, $500$, and $800$~K for n-
and p-type doping for $\varepsilon=0.00$, $0.05$ and $0.10$. The
data for $\varepsilon=0.00$ at $800$~K is split into 3 sets to
illustrate the 3-family behavior. The vertical dotted grey lines identify
the AGNRs belonging to the $N=3p+1$ family. }

\label{Fig.maxZT_300_500_800K} 
\end{figure}

We shall give a detailed discussion of the effect of strain on AGNR-$N$,
first with $N=3p+1$, then $N=3p$, and finally $N=3p+2$. Particular
attention will be paid to the $E_{g}$ variation in comparison with
$10kT$ as well as the changes in the $\theta_{{\rm e}}(E)$ due to
$\epsilon$ in determining the values of $G_{{\rm e}}$, $|S|$, $K_{{\rm e}}$,
$K_{{\rm ph}}$, and finally $ZT$. Unless otherwise stated, the temperature
is taken to be $800$~K for the discussion. For AGNR-$N$ with $N=3p+1$,
$G_{\text{e}}$ and $K_{\text{e}}$ values decrease with increasing
$\varepsilon$ for $N=4$ and $7$ (due mainly to the reduction in
$\theta_{{\rm e}}(E)$), but $G_{\text{e}}$ and $K_{\text{e}}$ increase
with increasing $\varepsilon$ due to bipolar transport for $N=10$.
For $|S|$, it remains essentially constant for $N=4$ and $7$, but
it decreases for $N=10$. We note that $|S|$ remains essentially
constant for $N=4$ and $7$ since it is proportional to $L_{1}/L_{0}$,
so the changes in $\theta_{\text{e}}(E)$ is somewhat suppressed when
the ratio is taken. The net outcome for changes in $G_{{\rm e}}$,
$|S|$, and $K_{{\rm e}}$, is shown in Fig.~\ref{Fig.maxZT_300_500_800K}(e)
where the $ZT$ value for AGNR-$(3p+1)$ decreases with increasing
$\varepsilon$.

Next we discuss the $N=3p$ family, where $\theta_{\text{e}}(E)$
and $E_{\text{g}}$ of AGNRs in the $N=3p$ family generally increases
with increasing $\varepsilon$. The increase in $\theta_{\text{e}}(E)$
causes $G_{\text{e}}$, $K_{\text{e}}$, and $|S|$ to increase with
$\varepsilon$. However, an increase in $E_{\text{g}}(\epsilon)$
does not significantly benefit $|S|$ because
even the smallest $E_{\text{g}}$ at $\varepsilon=0.00$ for all AGNR-$3p$
is already large enough to prevent bipolar transport. Overall, at
$T=800$~K, the increase in both $G_{\text{e}}$ and $K_{\text{e}}$
causes $ZT$ to increase with strain $\epsilon$ (see Fig.~\ref{Fig.maxZT_300_500_800K}(e)).

Since AGNR-$(3p+2)$ have the smallest $E_{\text{g}}$ among the three
families, therefore bipolar transport is present even at $\varepsilon=0.00$
that benefits $G_{{\rm e}}$ and $K_{{\rm e}}$ but not $|S|$. 
At $T=800$~K,
bipolar transport becomes dominant and causes $|S|$ to become small.
As $E_{\text{g}}$ increases with increasing $\varepsilon$, $G_{{\rm e}}$
and $K_{{\rm e}}$ decrease but $|S|$ increases since the monopolar
transport becomes more pronounced. The overall effect, however, is
to increase $ZT$ as shown in Fig.~\ref{Fig.maxZT_300_500_800K}(e).

We now examine the effect of $\varepsilon$ and temperature on $K_{\text{ph}}$
and its associated influence on $ZT$ of AGNRs. The effect of temperature
is also seen (see Fig.~\ref{Fig:N_vs_ABS_ZT_params}(d)) to increase
the thermal conductance due to phonon ($K_{{\rm ph}}$) from $300$~K
to $800$~K. However, at high temperatures, the effect of strain
on $K_{\text{ph}}$ for various $\epsilon$ could decrease the $K_{{\rm ph}}$
due an overall shifting down of the phonon frequencies as a result of weaker
interatomic interactions.\cite{Yeo2012} For example, at $T=800$~K,
a strain of $\epsilon=0.10$ could decrease $K_{{\rm ph}}$ by $\sim15~\%$
from that of the unstrained AGNRs. We note that the total thermal
conductance of AGNRs is dominated by $K_{\text{ph}}$ and not $K_{{\rm e}}$.

\Fig~\ref{Fig.maxZT_300_500_800K} shows the maximum $ZT$ values
attainable at $T=300$, $500$, and $800$~K for n- and p-type doping.
It can be seen that at higher temperatures, $ZT$ changes more sensitively
with $\varepsilon$ because the larger spread of $\frac{-\text{\ensuremath{\partial}}f}{\text{\ensuremath{\partial}}E}$
magnifies the changes to $\theta_{\text{e}}$ and $E_{\text{g}}$
due to $\varepsilon$. In general the $ZT$ value decreases with increasing
$N$ due to the faster increase in $K_{\text{ph}}$ as the AGNR-$N$
become wider. We observe that, under strain, $ZT$ increases for AGNR-$N$ belonging
to the $N=3p$ and $N=3p+2$ families, but decreases for the $N=3p+1$ family. 
It is interesting to note that
at $T=800$~K and $\epsilon=0.00$, bipolar transport becomes significant
for AGNRs with small $E_{\text{g}}$ such that the monotonic decrease
in $ZT$ values can be grouped according to three families, which
is largely due to the 3-family behavior\cite{Son2006a} exhibited by $E_{g}$.

\begin{figure}
\begin{centering}
\includegraphics[width=7.2cm]{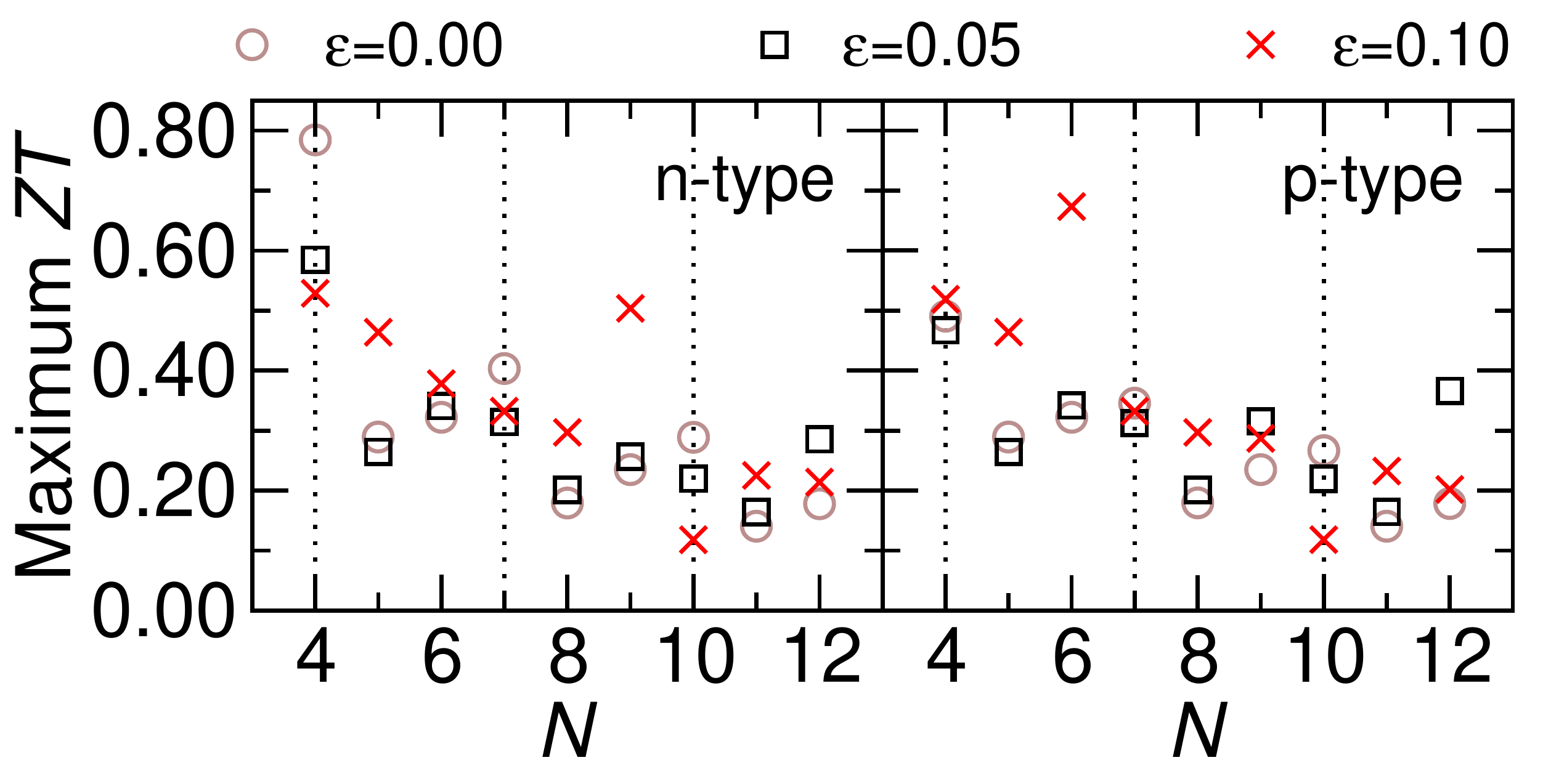} 
\par\end{centering}

\caption{Maximum attainable $ZT$ for AGNR-$N$ in the temperature range of
$200$ to $800$~K and doping concentration range of $-10^{14}$
to $10^{14}$~cm\textsuperscript{-2}. 
For $\varepsilon=0.10$, AGNR-$9$ and AGNR-$6$ have the largest $ZT$ values under n-type and p-type
doping, respectively. For AGNR-$9$ the maximum $ZT$ of 0.504 is achieved at $T=800$~K and a doping concentration of 
$-1.64\times10^{12}$~cm$^{-2}$; for AGNR-$6$, the maximum $ZT$ of 0.673 is achieved at $T=800$~K and a doping concentration of 
$3.30\times10^{13}$~cm$^{-2}$.
}

\label{Fig:maxZT} 
\end{figure}

\Fig~\ref{Fig:maxZT} shows the maximum $ZT$ value attainable for
AGNR-$N$ in the temperature range $T=200-800$~K and doping concentration
of $-10^{14}$ to $10^{14}$~cm\textsuperscript{-2}, giving an upper
limit to the $ZT$ value that can be achieved through strain engineering.
At the highest $\varepsilon=0.10$, $ZT$ increases by $13.3-114.3$\%
for AGNRs in the $N=3p$ family, and increases by $60.3-65.5$\% for
the $N=3p+2$ family. However, $ZT$ decreases by $3.9-59.1$\%
for the $N=3p+1$ family. 

\section{Conclusion}

We have calculated the thermoelectric figure of merit $ZT$ for AGNR-$N$
($N$ is the number of carbon dimer lines across the AGNR width) when
uniaxial tensile strain is applied along the main ribbon axis of the
AGNR. We have considered both n- and p-type doping concentrations
of up to $10^{14}$~cm\textsuperscript{-2}\ and a temperature range
of $200-800$~K. Using density-functional theory calculations, the
effect of $\varepsilon$ is found to improve $ZT$ for AGNR-$N$,
for $N=3p$ and $N=3p+2$. For the $N=3p$ family, this is due to
an increase in the electronic transmission around the valence and
conduction band edges. For the $N=3p+2$ family, it is due to an increase
in the band gap that reduces the unfavorable bipolar transport. Based
on first principles, we concluded that strain engineering provides
a possible route to improve the $ZT$ values of two families of AGNR-$N$.

\section*{Acknowledgements}

{We gratefully acknowledge the support of A{*}STAR Computational
Resource Center (A{*}CRC) of Singapore. }

\bibliography{references} 

\end{document}